\documentclass[english,prl,superscriptaddress,twocolumn]{revtex4-1}
\pdfoutput=1

\usepackage{graphicx}
\usepackage{textcomp}
\usepackage{amsmath}
\usepackage{amssymb}
\usepackage{pgf}
\usepackage{ulem}
\usepackage{amsfonts}
\usepackage{hyperref}
\graphicspath{{Figs/}{../Qpumping_dis/Notes/},{Figs/figW/}}

\newcommand{\ket}[1]{\left|#1\right\rangle}
\newcommand{\bra}[1]{\left\langle #1\right|}
\newcommand{\Ho}{\hat{H}}

\newcommand{\ud}{\mathrm{d}}
\newcommand{\nep}{\textrm{e}}
\newcommand{\mean}[1]{\left\langle #1\right\rangle}

\newcommand{\opc}[1]{{\hat{c}^{\phantom \dagger}}_{#1}}
\newcommand{\opcdag}[1]{{\hat{c}^{\dagger}}_{#1}}

\newcommand{\calE}{\mathcal{E}}
\newcommand{\calN}{\mathcal{N}}

\newcommand{\IPR}{\mathrm{IPR}}

\begin{document}
%
\title{Localization, topology and quantized transport in disordered Floquet systems}

\vspace{1cm}
\author{Matteo M. Wauters}
\affiliation{SISSA, Via Bonomea 265, I-34136 Trieste, Italy}
\author{Angelo Russomanno}
\affiliation{International Centre for Theoretical Physics (ICTP), P.O.Box 586, I-34014 Trieste, Italy}
\author{Roberta Citro}
 \affiliation{Dipartimento di Fisica ``E.R. Caianiello'', Universit\`a di Salerno and Spin-CNR, Via Giovanni Paolo II, 132, I-84084 Fisciano
  (Sa), Italy}
\author{Giuseppe E. Santoro}
\affiliation{SISSA, Via Bonomea 265, I-34136 Trieste, Italy}
\affiliation{International Centre for Theoretical Physics (ICTP), P.O.Box 586, I-34014 Trieste, Italy}
\affiliation{CNR-IOM Democritos National Simulation Center, Via Bonomea 265, I-34136 Trieste, Italy}

\author{Lorenzo Privitera}
\affiliation{Institute for Theoretical Physics, University of W\"urzburg, 97074 W\"urzburg, Germany}

\begin{abstract}
We investigate the effects of disorder on a periodically-driven one-dimensional model displaying quantized topological transport. 
We show that,  while instantaneous eigenstates are necessarily Anderson localized, the periodic driving plays a fundamental role in 
delocalizing Floquet states over the whole system, henceforth allowing for a steady-state nearly-quantized current.
Remarkably, this is linked to a localization/delocalization transition in the Floquet states at strong disorder, which occurs for periodic driving corresponding to a non-trivial loop in the parameter space.
As a consequence, the Floquet spectrum becomes continuous in the delocalized phase, in contrast with a pure-point instantaneous spectrum.
\end{abstract}

\maketitle
{\em Introduction}.
Thouless pumping~\cite{thouless1982quantized,Niu_JPA84} 
provides one of the simplest manifestations of topology in quantum systems, and has attracted a lot of recent interest, both 
theoretically~\cite{Chern_PRB07, kitagawa2010topological, Qin_PL16, Lindner_PRX16,  Privitera_PRL18, Moore_PRL18,Nakagawa_PRB18,Haug_arxiv19,Arceci_Arx19} 
and experimentally~\cite{Lu_NatPhot14,Lohse_Nphys16, Nakajima_Nphys16, Lohse_Nat18,Ma_PRL18,Hotzen_Arx19}.
Since the seminal works by Thouless and Niu~\cite{thouless1982quantized,Niu_JPA84}, it is argued that the quantization of the pumped charge is 
robust against weak disorder, but a clear characterization of the localization properties of the relevant states, and the breakdown of quantized transport 
at strong disorder, is still missing. 

Thouless pumping is also the first example of a topological phase emerging in a periodically driven system with no static analogue. Such phases have been the subject of many recent 
proposals~\cite{kitagawa2010topological, rudner2013anomalous, Potter_PRX16, Else_PRB16, Lindner_PRX16, Budich_PRL17, Moore_PRL18,Lindner_PRB18}.
%
In this respect, understanding the role of disorder has a twofold purpose: 
on one hand, it is important to understand the robustness to disorder of the topology of driven systems~\cite{Hotzen_Arx19,Shtanko_PRL18} {\em per se};
on the other hand, localization properties in the topological phase are relevant for the possibility of stabilizing topological pumping in {\em interacting} systems~\cite{Lindner_PRX17,Gulden_arxiv19}
by means of many-body localization~\cite{Bloch1,Harper_arxiv19}.

Restricting ourselves to the non-interacting case, a puzzling aspect regards the nature of the Floquet states at long time. 
While quantized transport over a single period of the driving is expected at small disorder\cite{Niu_JPA84}, its robustness over many driving cycles is not trivial, since it would imply the existence of {\em extended} Floquet states. 
But in the adiabatic limit, where charge is strictly quantized, Floquet states for a generic driving should coincide
with the Hamiltonian eigenstates, which are Anderson localized in 1d. 
So, how can Thouless pumping in Anderson insulators be stable in the long-time limit?
Previous studies of periodically-driven 1d  Anderson insulators in the low-frequency regime~\cite{Hatami_PRE16, Agarwal_PRB17} have found 
a generic increase of the localization length of Floquet states compared to the static case, without any evidence of truly extended states nor a clear link 
between the localization properties and topology.

In this work, we address these questions by inquiring the effects of disorder on Thouless pumping from the point of view of Floquet theory.
We focus on the finite-size scaling of the localization length of Floquet states, the long-time dynamics and the winding of Floquet quasienergies, 
and show that Thouless pumping is associated to extended Floquet states.
Remarkably, as disorder increases these states undergo a true delocalization/localization transition at a critical disorder strength $W_c$, 
which reflects itself in the breakdown of quantized transport. 
Crucially, topology plays a fundamental role in the existence of such extended states and on the character of the phase transition, as we prove by explicit comparison with the case of a trivial adiabatic driving protocol. 

{\em Model}. 
We consider a disordered version of  the driven Rice-Mele model~\cite{Rice_PRL82}.
For a system of spinless fermions on a chain of $L=2N$ sites, with $\opcdag{j}$ creating a fermion on the $j-$th site, the Hamiltonian reads
\begin{equation} \label{eq:RM-model}
\begin{split}
\hat{H}(t) =&
- \sum_{j=1}^N \left[ J_1(t) \, \opcdag{2j-1}\opc{2j} + J_2(t) \, \opcdag{2j}\opc{2j+1} + {\rm H.c.} \right] \\
&- \sum_{l=1}^{L} \left[(-1)^{l}\Delta(t) +W \zeta_l  \right] \opcdag{l}\opc{l}\;.
\end{split}
\end{equation}
Here $J_{1(2)}(t)$ and $\Delta(t)$ describe hopping amplitudes and on-site energies for the clean model, 
while $W\zeta_l$ describes the on-site disorder of strength $W$, with $\zeta_l\in [-\frac{1}{2},\frac{1}{2}]$ uniformly distributed random numbers. 
We assume periodic boundary conditions. 
In absence of disorder, $W=0$, and for generic $J_{1(2)}$ and $\Delta$, the instantaneous spectrum is split in two bands, separated by a gap. Thus, at half filling, the charge pumped in one period is equal in the adiabatic limit to the  Chern number of the occupied band~\cite{thouless1982quantized}. This integer is different from $0$ when the driving is topologically non-trivial, e.g. when the path in the space $(J_1-J_2,\Delta)$ encloses the gapless point $(0,0)$.  

To characterize the topological phase we compute the infinite-time average of the pumped charge~\cite{Avron_JPA99,Privitera_PRL18}
\begin{equation}\label{eq:Qd}
\overline{Q}= \frac{1}{L} \lim_{M\to \infty}\frac{1}{M} \int_0^{M\tau} \ud t \langle \Psi(t)|\hat{\mathrm J}(t)|\Psi(t)\rangle \ , 
\end{equation}
where $\hat{\mathrm J}(t)$ is the total current operator, $\tau=2\pi/\omega$ is the driving period, and the system is initially prepared in the $N$-particle 
ground state $|\Psi_0\rangle$ of $\Ho(t=0)$.

Since the Hamiltonian is time periodic, we can exploit the Floquet representation~\cite{Grifoni1998driven} of the evolution operator 
$\hat{U}(t,0)=\sum_\nu \nep^{-i \calE_\nu t/\hbar} \ket{\Phi_\nu(t)} \bra{\Phi_\nu(0)}$, 
where $\ket{\Phi_\nu(t)}=\ket{\Phi_\nu(t+\tau)}$ are $N$-particle Floquet modes, while $\calE_\nu$ are the many-body quasienergies.
$\overline{Q}$ can be computed directly in the Floquet diagonal ensemble~\cite{Russomanno_PRL12,Privitera_PRL18} 
\begin{equation}\label{eq:Qd_floq}
\overline{Q}=Q_d=\frac{1}{L} \sum_{\nu} \calN_\nu \int_0^\tau \! \ud t \; \langle \Phi_\nu(t)|\hat{\mathrm J}(t)|\Phi_\nu(t)\rangle \;, 
\end{equation}
where ${\calN}_\nu =|\langle \Psi_0|\Phi_\nu (0) \rangle|^2$ is the occupation number of the $\nu$-th Floquet state.
For non-interacting fermions, it suffices to know the single-particle (SP) Floquet states $\ket{\phi_\alpha(t)}$ and their occupation number $n_\alpha $ to explicitly calculate the 
diagonal pumped charge~\cite{Avron_JPA99,Privitera_PRL18}. 

\begin{figure}[t]
\begin{center}
\includegraphics[scale=0.60]{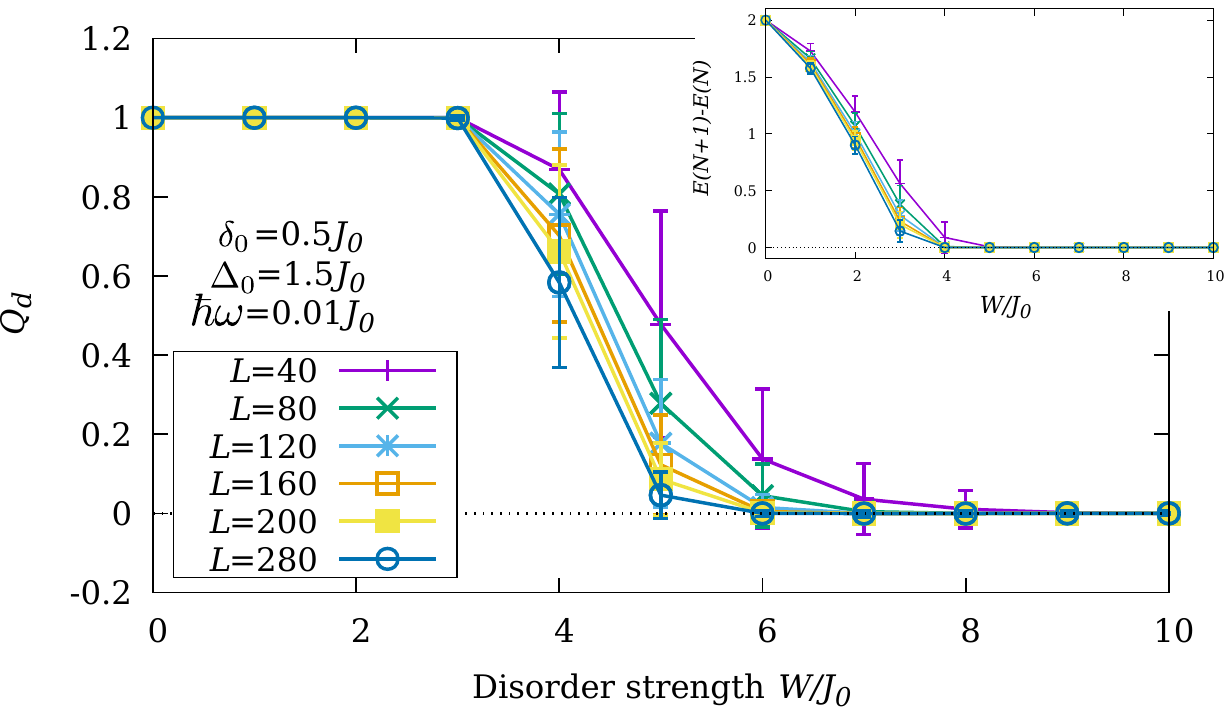}
\end{center}
\caption{Disorder average of the diagonal pumped charge plotted against disorder strength. 
The transition between the quantized charge regime and the trivial one $Q_d=0$ is clearly linked to the closing of the minimum energy gap due to the disorder, shown in the inset.
}\label{fig:Qd_dis}
\end{figure}

{\em Results}.
Figure~\ref{fig:Qd_dis} shows the disorder average $[Q_d]_{av}$ as a function of the disorder strength $W$. 
Observe that topological pumping persists for sufficiently small $W\lesssim 3J_0$. 
The regime of large $W\gtrsim 8 J_0$ is also rather clear: $Q_d=0$.
The intermediate region $W/J_0\approx 4$ shows large sample-to-sample fluctuations: the inset shows a correlation between the drop of $[Q_d]_{av}$
and the closing of the minimal many-body instantaneous gap $\Lambda_N \equiv \min_{t\in [0,\tau]} [E_{N+1}(t)-E_N(t)]$, where $E_N(t)$
is the $N$-particle ground state energy at time $t$.   

A natural question arises: if a disordered one-dimensional system shows Anderson-localized instantaneous
energy eigenstates and a pure-point spectrum~\cite{Abrahams_PRL79}, is it actually able to transport charge?
This fact is puzzling, considering that the charge transport is  precisely quantized in the {\em adiabatic} limit, where Floquet states should coincide, apart from a phase, with the instantaneous Hamiltonian eigenstates.  
In the following, we will show that the crucial ingredient behind topological pumping is that a significant fraction of the SP Floquet states
remain {\em delocalized} even for very low frequency.
The key point is that the dynamics is adiabatic only at many-body level, but not at the SP one, where driving-induced mixing of localized states occurs~\cite{Agarwal_PRB17,Hatami_PRE16}. 

To address the issue of localization/delocalization of states, we analyze the real-space inverse participation ratio (IPR)~\cite{Edwards_JPC72} of the single-particle 
Floquet modes $|\phi_{\alpha}(0)\rangle$,  
$\IPR_\alpha = \sum_l |\langle l \ket{\phi_\alpha(0)}|^4 $, with $\ket{l}=\opcdag{l} |0\rangle$ being a particle localized at site $l$. 
\begin{figure*}[t]
\begin{center}
\includegraphics[scale=0.35]{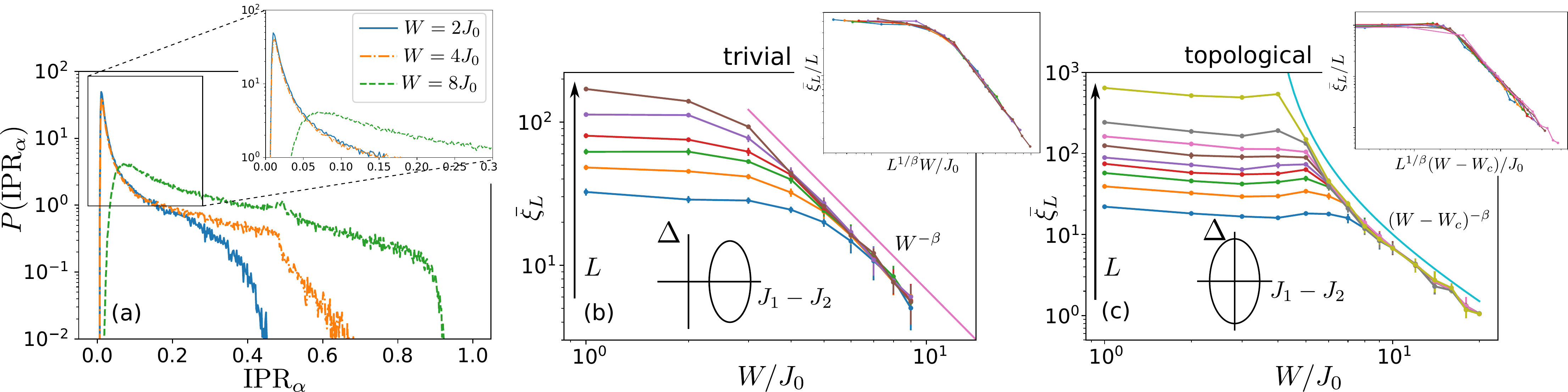}
\caption{(a): Disorder averaged $\IPR$ distribution of SP Floquet states for several values of $W$ and $L=280$.
The inset shows the sharp peaks almost superimposed at small IPR for $W=2J_0$ and $W=4J_0$.
(b): Characteristic localization length $\overline{\xi}_L(W)$ as a function of disorder, for a trivial driving cycle ($J_1=4J_0+\delta_0\sin(\omega t)$). 
The inset shows the collapse obtained by $\overline{\xi}_L \to  \overline{\xi}_L/L$ and $W \to W L^{1/\beta}$, with $\beta\simeq 2.5$.
(c): Characteristic localization length when the system exhibits topological transport ($J_1=J_0+\delta_0\sin(\omega t)$). 
The inset shows the collapse $\overline{\xi}_L \to  \overline{\xi}_L/L$ and $W \to (W-W_c)L^{1/\beta}$, with $W_c \simeq 3.5 J_0$ and $\beta\simeq 2$.
The parameters used in the simulation are $\delta_0=0.5J_0$, $\Delta_0=1.5J_0$, $\hbar\omega=0.01J_0$.}
\label{fig:IPR_dist_dis}
\end{center}
\end{figure*}
For a finite system $\IPR_\alpha \in [L^{-1},1]$, where $\IPR_\alpha\sim L^{-1}$ signals a completely delocalized (plane-wave-like) state, 
while $\IPR_\alpha=1$ corresponds to a perfect localization on a single site.
Figure~\ref{fig:IPR_dist_dis}(a) shows the distribution of IPRs of Floquet states for three values of the disorder strength $W$.
Notice the presence of a very sharp peak in the IPR distribution which we find to scale as $\IPR_\alpha\sim L^{-1}$ for $W/J_0=2$ and $4$, 
suggesting that the {\em mode} of the $\IPR$ distribution corresponds to extended states~\cite{nota1}.
We find, however, that very similar distributions (not shown) would emerge --- for the same disorder strength --- when the driving protocol is topologically {\em trivial}.

To better analyze the size dependence of the $\IPR$ peak, and its correlation with the topology of the driving, 
we estimate a characteristic localization length for a chain of size $L$ from the inverse of the peak's position in the $\IPR$ distribution (inverse of the mode), 
$\overline{\xi}_L(W)=({\mathrm{argmax}}(P(\IPR_\alpha)))^{-1}$. 
Fig.~\ref{fig:IPR_dist_dis}(b) and (c) show the size-scaling of $\overline{\xi}_L(W)$ for a trivial and topological driving, respectively. 
When the driving is trivial, our data suggest that $\overline{\xi}_L(W)$ scales as $W^{-\beta}$ with $\beta \simeq 2.5$ for large $W$, 
see Fig.~\ref{fig:IPR_dist_dis}(b), while saturates to the system size $\sim L$ when $W$ is small. 
Hence we can extract a crossover disorder strength $W^* \sim L^{-1/\beta}$ separating these two regimes, vanishing in the thermodynamic limit: here
truly extended Floquet states appear only at zero disorder.
By rescaling the data, we see a very good collapse of  $\overline{\xi}_L(W)/L$ versus $L^{1/\beta}W$ [Fig.~\ref{fig:IPR_dist_dis}(b)--inset]. 
On the other hand, when the driving is topological the same phenomenology holds with a {\em finite critical disorder strength} $W_c$ [Fig.~\ref{fig:IPR_dist_dis} (c)]: 
	For $W>W_c\simeq 3.5 J_0$, we observe that $\overline{\xi}_L(W)\sim (W-W_c)^{-\beta}$, with $\beta \simeq 2$, while again the localization length saturates to $L$ when $W<W_c$, thus indicating the presence of an actual localization/delocalization phase transition. The critical exponent is in good agreement with bosonization calculations~\cite{Citro_unpub}, while the value of $W_c$ extracted by our scaling analysis is compatible with the breaking of quantization in Fig.~\ref{fig:Qd_dis}. 

\begin{figure}[t]
\begin{center}
\includegraphics[scale=0.25]{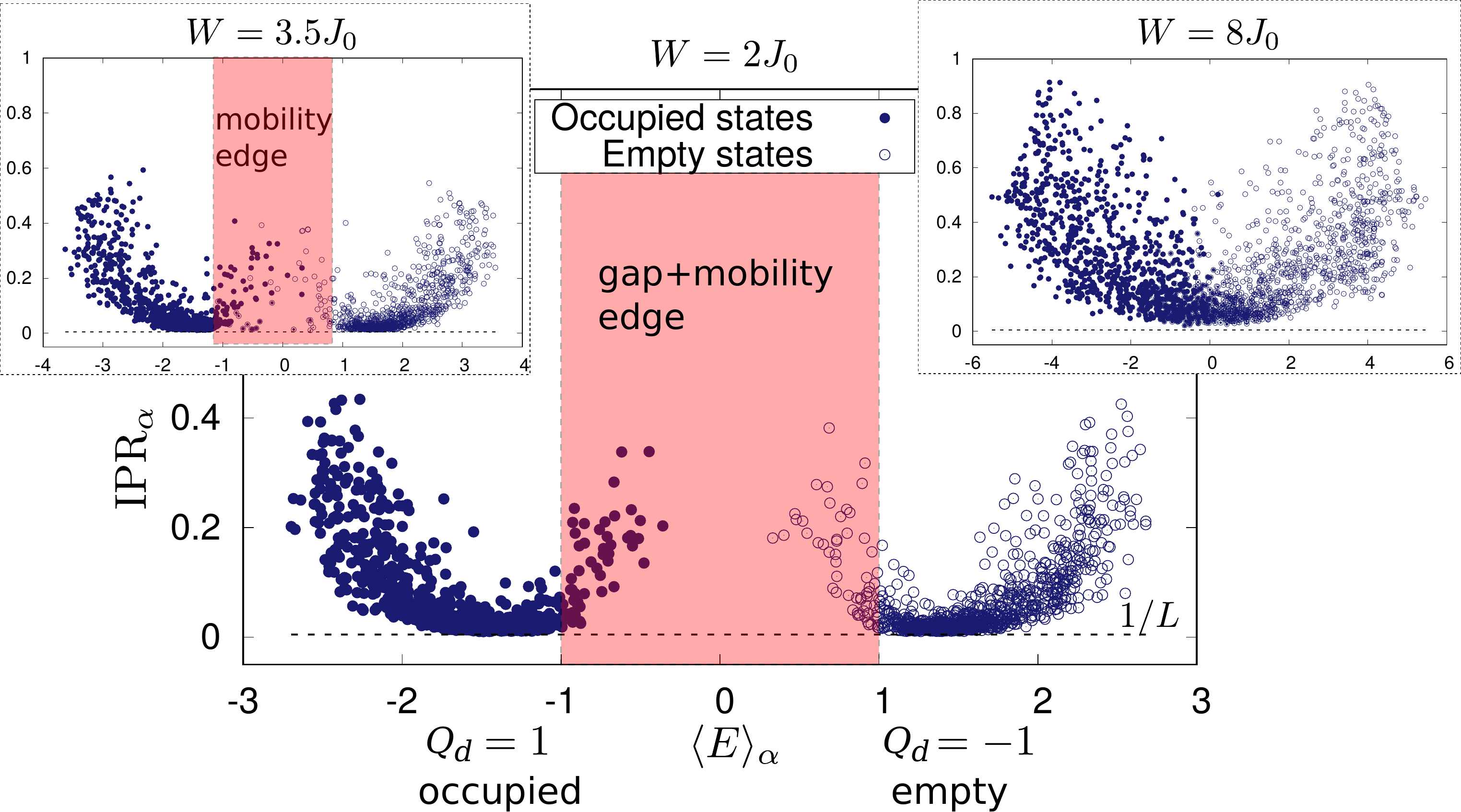}
\caption{$\IPR_\alpha$ vs the time-averaged energy of the corresponding Floquet state $\langle E \rangle_\alpha$.
The dashed lines indicates the value $1/L$ associated to  extended states; 
the data refer to several realizations of a chain with 200 sites and disorder strength $W=2J_0$.
In the insets the same data are shown for larger disorder, 
when delocalized states are separated only by a mobility edge ($W=3.5J_0$) 
and when there is a single band of localized Floquet states ($W=8J_0$).
$Q_d=\pm 1$ is the charge transported when a single band, the lower or the higher one, is completely filled. } \label{fig:IPRvsE}
\end{center}
\end{figure}

To better understand the mechanism behind the delocalization/localization transition,
we study the relation between the time-averaged energy of the SP Floquet states $\mean{E}_\alpha=\frac{1}{\tau}\int_0^\tau\ud\tau\bra{\phi_\alpha(t)}\hat{H}(t)\ket{\phi_\alpha(t)}$  
and the corresponding $\IPR_\alpha$ (Fig.~\ref{fig:IPRvsE}).
For weak disorder, extended states carrying charge in the positive (negative) direction lays in the middle of the lower (higher) band, while localized ones stay closer to the edges. 
	Hence quantized transport is protected both by an energy gap and a mobility edge.
	As disorder increases (left inset with $W=3.5J_0$), the energy gap is closed by localized states but there is still a mobility edge protecting the extended states and transport.
	For even larger $W$ the two bands merge in a single one where extended states transporting opposite charges gradually hybridize and give localized states,
	(inset with $W=8J_0$),
	and quantized transport breaks down.

This phenomenology is similar to what happens in integer quantum Hall effect (IQHE) in 2D systems, 
where there {\em must be} spectral regions of extended states~\cite{Prange_book90,Prodan_PRL2010,Prodan_JPA2011},
in order to have a non-zero quantized  transverse conductivity.
Also the exponent $\beta \simeq 2$ found for $W>W_c$, when the driving is topological, is in good agreement with a similar scaling analysis performed on the density of extended states in IQHE in a disordered sample~\cite{Yang_PRL96}.
The parallelism between the physics of clean 1D topological charge pumping and 2D integer quantum Hall effect is well established~\cite{Niu_JPA84,Avron_PT03,Martin_PRX17},
but at the best of our knowledge this is the first time where a localization/delocalization transition in a driven Anderson insulator is clearly associated to the topology of the clean limit,
as it happens in IQHE~\cite{Prodan_PRL2010,Prodan_JPA2011}.

{\em Winding of quasienergies}.
\begin{figure*}[t]
\begin{center}
\includegraphics[scale=0.13]{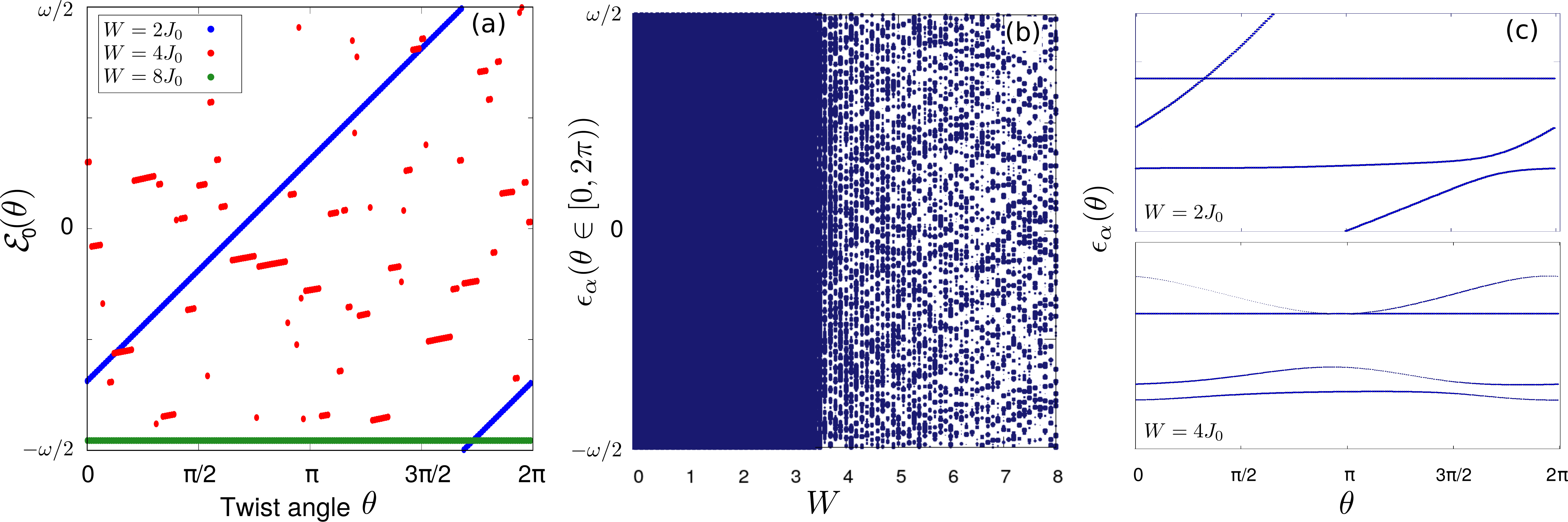}
\caption{(a) Quasienergy $\calE_0(\theta)$ of the many-body Floquet state with lowest energy. The winding is well defined only for $W=2J_0$ and $W=8J_0$, when $\ket{\Psi_0}$ has a non vanishing projection on a single MB Floquet state.
(b): SP quasienergy spectrum for all possible angles $\theta$ as a function of the disorder.
(c): typical behaviors of SP $\epsilon_\alpha(\theta)$. When $W<W_c$ most of the states are extended and sensitive to the boundary condition (upper panel),
while for $W > W_c$ many quasi-energies do not contribute to the winding, being periodic in  $\theta$ (lower panel). 
All data correspond to the same choice of the random variables $\lbrace \zeta_j \rbrace$ for $L=80$ sites. The size of the dots is proportional to $n_\alpha(\theta)$.}\label{fig:winding2}
\end{center}
\end{figure*}
In a clean system, quantized pumping corresponds to a non-trivial winding  of the quasienergy of the occupied Floquet bands in $k$-space~\cite{Shih_PRB94,Ferrari_IJMP98,Avron_JPA99,kitagawa2010topological,Privitera_PRL18}. 
When translational invariance is broken, a common procedure is to introduce a phase twist $\theta \in [0,2\pi)$ between site 1 and site $L$ and then take the average of $Q_d$ over $\theta$~\cite{Niu_JPA84,kitagawa2010topological}. 
This operation is justified because when the state projector is exponentially localized,
the dependence of observables on the twisted boundary decays exponentially with $L$~\cite{Watanabe_PRB18}.
Hence we can write
\begin{equation}\label{eq:Qd_winding}
Q_d=\int_0^{2\pi} \frac{\ud \theta}{2\pi} Q_d(\theta)= \tau \sum_\nu \int_{0}^{2\pi} \frac{\ud \theta}{2\pi} \partial_\theta \calE_{\nu}(\theta) \calN_\nu(\theta) \;.
\end{equation}
Here $\calE_{\nu}=\epsilon_{\alpha_1}+\dots + \epsilon_{\alpha_N}$ is the $N-$particle quasienergy associated to the Floquet state $\ket{\Phi_\nu}$ given by a Slater determinant of the SP states $\ket{\phi_{\alpha_1}},\dots,\ket{\phi_{\alpha_N}}$;
$\calN_\nu=|\langle \Psi_0\ket{\Phi_\nu}|^2$ is the occupation number. In this context, the winding number is the number of times that $\calE_{\nu}(\theta)$ wraps around the first Floquet-Brillouin zone as $\theta$ goes from 0 to $2\pi$.
Besides non-adiabatic corrections that depend on the initial state $\ket{\Psi_0}$~\cite{Privitera_PRL18,Wauters_PRB18},
 $Q_d$ is quantized when a single many-body Floquet state is occupied, e.g. $\calN_\nu \simeq \delta_{0,\nu}$ independently of $\theta$, and that state has a non trivial winding number. 
Henceforth we focus on the Floquet state with the lowest initial energy  $\ket{\Phi_0(\theta)}$, computed as the Slater determinant of the $N$ SP Floquet states with highest projection on the ground state in which the state is initially prepared.

We report in Fig.~\ref{fig:winding2}(a) $\calE_0(\theta)$ in the first Floquet-Brillouin zone for three different disorder strength $W/J_0=2$, 4 and 8, which are respectively  below, close to the transition value and above it.
In Fig.~\ref{fig:winding2}(b) the SP quasienergy spectrum is plotted with respect to $W$ as $\theta$ spans the interval $ \in [0,2\pi)$,
while Fig.~\ref{fig:winding2}(c) shows some details of $\epsilon_\alpha(\theta)$ that help to understand the localization transition. 
A localized state is characterized by a quasienergy $\epsilon_\alpha$ {\em periodic} in $\theta$, while extended ones with positive winding satisfy the relation 
$\epsilon_\alpha(2\pi)=\epsilon_{\alpha+1}(0)$.
Hence we distinguish between three situations. 
\begin{itemize}	
\item[-] $W<W_c$: 
$\ket{\Phi_0}$ coincides essentially with $|\Psi_0\rangle$  ($\calN_\nu\simeq \delta_{0,\nu}$) and has winding number equal to 1, blue line in Fig.~\ref{fig:winding2}(a).
The SP quasienergy spectrum is continuous in $\theta$ and there are no gaps in the Floquet-Brillouine zone, Fig.~\ref{fig:winding2}(b).
Most of the SP states feel the twist at the boundary, obey $\epsilon_\alpha(2\pi)=\epsilon_{\alpha+1}(0)$ and contribute to the winding of $\calE(\theta)$ (upper panel of Fig.~\ref{fig:winding2}(c)).

\item[-] $W\simeq W_c$:  the  initial ground state $\ket{\Psi_0}$ has relevant projections over many Floquet states and thus Eq.~\eqref{eq:Qd_winding} requires all combinations of $N$ SP Floquet states.
Gaps start to appear in the SP quasienergy spectrum(Fig.~\ref{fig:winding2}(b)), and the occupation number itself depends non-trivially on $\theta$.
$\calE_0(\theta)$ is discontinuous $\theta$ [red dots in Fig.~\ref{fig:winding2}(a)].
The SP Floquet states with opposite transported charge start to be mixed in pairs of localized states, with quasienergies periodic in $\theta$ (lower panel of Fig.~\ref{fig:winding2}(c)).

\item[-] $W>W_c$: both SP Floquet states and Hamiltonian eigenstates are strongly localized and there is no current. 
Again $\calN_\nu=\delta_{0,\nu}$, but the winding number is trivial (green line in Fig.~\ref{fig:winding2}(a))	because SP quasienergy spectrum has only a pure-point contribution and localization makes the system insensitive to the boundary twist.
\end{itemize}

{\em Conclusions}. 
We analyzed in detail the steady-state current flowing in a one dimensional Floquet-Anderson insulator: the topological periodic driving mixes localized Hamiltonian eigenstates to give extended Floquet modes.
Delocalization makes quantized pumping robust, until extended Floquet states with opposite winding coalesce for large disorder.
Even though the physics of quantum pumping in clean systems is the same of 2D IQHE, this analogy is not trivial in the presence of disorder, since the 1D periodically driven chain would be mapped in an {\em extremely anisotropic} disordered 2D model. 

Finally, a subtle point emerges in the adiabatic limit $\omega\to 0$. 
In a truly adiabatic evolution the Floquet states would coincide with the Hamiltonian eigenstates, thus being localized, at least when the disorder-induced SP level crossings are actually {\em avoided crossings}~\cite{nota2}.
However the quantization of the pumped charge requires both the adiabatic {\em and} thermodynamic limit, but when $L\to \infty$ the spectrum becomes {\em dense} and the driving mixes SP levels.
On the opposite, when one takes the adiabatic limit in a finite system, SP Floquet states are identical to Hamiltonian eigenstates, modulo a phase factor, and are localized.
Quantized pumping still works because disorder induces resonances in the spectrum at any energy scale, allowing for large distance tunneling~\cite{Khemani_NatPhys15,Ippoliti_arx19}.

The interplay between disorder, topology and possibly interaction in Floquet systems can be investigated in cold atom experiments, where a  disordered or quasi-periodic potential can be easily engineered \cite{Bordia_NatPhys17}.

\acknowledgements{We wish to thank M.~Dalmonte, R.~Fazio, N.~Lindner, A.~Michelangeli, B.~Trauzettel, V.~Ros and J.~Asboth for valuable discussions. }


\begin{thebibliography}{52}%
\makeatletter
\providecommand \@ifxundefined [1]{%
 \@ifx{#1\undefined}
}%
\providecommand \@ifnum [1]{%
 \ifnum #1\expandafter \@firstoftwo
 \else \expandafter \@secondoftwo
 \fi
}%
\providecommand \@ifx [1]{%
 \ifx #1\expandafter \@firstoftwo
 \else \expandafter \@secondoftwo
 \fi
}%
\providecommand \natexlab [1]{#1}%
\providecommand \enquote  [1]{``#1''}%
\providecommand \bibnamefont  [1]{#1}%
\providecommand \bibfnamefont [1]{#1}%
\providecommand \citenamefont [1]{#1}%
\providecommand \href@noop [0]{\@secondoftwo}%
\providecommand \href [0]{\begingroup \@sanitize@url \@href}%
\providecommand \@href[1]{\@@startlink{#1}\@@href}%
\providecommand \@@href[1]{\endgroup#1\@@endlink}%
\providecommand \@sanitize@url [0]{\catcode `\\12\catcode `\$12\catcode
  `\&12\catcode `\#12\catcode `\^12\catcode `\_12\catcode `\%12\relax}%
\providecommand \@@startlink[1]{}%
\providecommand \@@endlink[0]{}%
\providecommand \url  [0]{\begingroup\@sanitize@url \@url }%
\providecommand \@url [1]{\endgroup\@href {#1}{\urlprefix }}%
\providecommand \urlprefix  [0]{URL }%
\providecommand \Eprint [0]{\href }%
\providecommand \doibase [0]{http://dx.doi.org/}%
\providecommand \selectlanguage [0]{\@gobble}%
\providecommand \bibinfo  [0]{\@secondoftwo}%
\providecommand \bibfield  [0]{\@secondoftwo}%
\providecommand \translation [1]{[#1]}%
\providecommand \BibitemOpen [0]{}%
\providecommand \bibitemStop [0]{}%
\providecommand \bibitemNoStop [0]{.\EOS\space}%
\providecommand \EOS [0]{\spacefactor3000\relax}%
\providecommand \BibitemShut  [1]{\csname bibitem#1\endcsname}%
\let\auto@bib@innerbib\@empty
\bibitem [{\citenamefont {Thouless}\ \emph {et~al.}(1982)\citenamefont
  {Thouless}, \citenamefont {Kohmoto}, \citenamefont {Nightingale},\ and\
  \citenamefont {Den~Nijs}}]{thouless1982quantized}%
  \BibitemOpen
  \bibfield  {author} {\bibinfo {author} {\bibfnamefont {D.}~\bibnamefont
  {Thouless}}, \bibinfo {author} {\bibfnamefont {M.}~\bibnamefont {Kohmoto}},
  \bibinfo {author} {\bibfnamefont {M.}~\bibnamefont {Nightingale}}, \ and\
  \bibinfo {author} {\bibfnamefont {M.}~\bibnamefont {Den~Nijs}},\ }\href@noop
  {} {\bibfield  {journal} {\bibinfo  {journal} {Phys. Rev. Lett.}\ }\textbf
  {\bibinfo {volume} {49}},\ \bibinfo {pages} {405} (\bibinfo {year}
  {1982})}\BibitemShut {NoStop}%
\bibitem [{\citenamefont {Niu}\ and\ \citenamefont
  {Thouless}(1984)}]{Niu_JPA84}%
  \BibitemOpen
  \bibfield  {author} {\bibinfo {author} {\bibfnamefont {Q.}~\bibnamefont
  {Niu}}\ and\ \bibinfo {author} {\bibfnamefont {D.}~\bibnamefont {Thouless}},\
  }\href@noop {} {\bibfield  {journal} {\bibinfo  {journal} {J. Phys. A-Math.
  Gen.}\ }\textbf {\bibinfo {volume} {17}},\ \bibinfo {pages} {2453} (\bibinfo
  {year} {1984})}\BibitemShut {NoStop}%
\bibitem [{\citenamefont {Chern}\ \emph {et~al.}(2007)\citenamefont {Chern},
  \citenamefont {Onoda}, \citenamefont {Murakami},\ and\ \citenamefont
  {Nagaosa}}]{Chern_PRB07}%
  \BibitemOpen
  \bibfield  {author} {\bibinfo {author} {\bibfnamefont {C.-H.}\ \bibnamefont
  {Chern}}, \bibinfo {author} {\bibfnamefont {S.}~\bibnamefont {Onoda}},
  \bibinfo {author} {\bibfnamefont {S.}~\bibnamefont {Murakami}}, \ and\
  \bibinfo {author} {\bibfnamefont {N.}~\bibnamefont {Nagaosa}},\ }\href
  {\doibase 10.1103/PhysRevB.76.035334} {\bibfield  {journal} {\bibinfo
  {journal} {Phys. Rev. B}\ }\textbf {\bibinfo {volume} {76}},\ \bibinfo
  {pages} {035334} (\bibinfo {year} {2007})}\BibitemShut {NoStop}%
\bibitem [{\citenamefont {Kitagawa}\ \emph {et~al.}(2010)\citenamefont
  {Kitagawa}, \citenamefont {Berg}, \citenamefont {Rudner},\ and\ \citenamefont
  {Demler}}]{kitagawa2010topological}%
  \BibitemOpen
  \bibfield  {author} {\bibinfo {author} {\bibfnamefont {T.}~\bibnamefont
  {Kitagawa}}, \bibinfo {author} {\bibfnamefont {E.}~\bibnamefont {Berg}},
  \bibinfo {author} {\bibfnamefont {M.}~\bibnamefont {Rudner}}, \ and\ \bibinfo
  {author} {\bibfnamefont {E.}~\bibnamefont {Demler}},\ }\href@noop {}
  {\bibfield  {journal} {\bibinfo  {journal} {Phys. Rev. B}\ }\textbf {\bibinfo
  {volume} {82}},\ \bibinfo {pages} {235114} (\bibinfo {year}
  {2010})}\BibitemShut {NoStop}%
\bibitem [{\citenamefont {Qin}\ and\ \citenamefont {Guo}(2016)}]{Qin_PL16}%
  \BibitemOpen
  \bibfield  {author} {\bibinfo {author} {\bibfnamefont {J.}~\bibnamefont
  {Qin}}\ and\ \bibinfo {author} {\bibfnamefont {H.}~\bibnamefont {Guo}},\
  }\href {\doibase https://doi.org/10.1016/j.physleta.2016.05.014} {\bibfield
  {journal} {\bibinfo  {journal} {Physics Letters A}\ }\textbf {\bibinfo
  {volume} {380}},\ \bibinfo {pages} {2317 } (\bibinfo {year}
  {2016})}\BibitemShut {NoStop}%
\bibitem [{\citenamefont {Titum}\ \emph {et~al.}(2016)\citenamefont {Titum},
  \citenamefont {Berg}, \citenamefont {Rudner}, \citenamefont {Refael},\ and\
  \citenamefont {Lindner}}]{Lindner_PRX16}%
  \BibitemOpen
  \bibfield  {author} {\bibinfo {author} {\bibfnamefont {P.}~\bibnamefont
  {Titum}}, \bibinfo {author} {\bibfnamefont {E.}~\bibnamefont {Berg}},
  \bibinfo {author} {\bibfnamefont {M.~S.}\ \bibnamefont {Rudner}}, \bibinfo
  {author} {\bibfnamefont {G.}~\bibnamefont {Refael}}, \ and\ \bibinfo {author}
  {\bibfnamefont {N.~H.}\ \bibnamefont {Lindner}},\ }\href {\doibase
  10.1103/PhysRevX.6.021013} {\bibfield  {journal} {\bibinfo  {journal} {Phys.
  Rev. X}\ }\textbf {\bibinfo {volume} {6}},\ \bibinfo {pages} {021013}
  (\bibinfo {year} {2016})}\BibitemShut {NoStop}%
\bibitem [{\citenamefont {Privitera}\ \emph {et~al.}(2018)\citenamefont
  {Privitera}, \citenamefont {Russomanno}, \citenamefont {Citro},\ and\
  \citenamefont {Santoro}}]{Privitera_PRL18}%
  \BibitemOpen
  \bibfield  {author} {\bibinfo {author} {\bibfnamefont {L.}~\bibnamefont
  {Privitera}}, \bibinfo {author} {\bibfnamefont {A.}~\bibnamefont
  {Russomanno}}, \bibinfo {author} {\bibfnamefont {R.}~\bibnamefont {Citro}}, \
  and\ \bibinfo {author} {\bibfnamefont {G.~E.}\ \bibnamefont {Santoro}},\
  }\href {\doibase 10.1103/PhysRevLett.120.106601} {\bibfield  {journal}
  {\bibinfo  {journal} {Phys. Rev. Lett.}\ }\textbf {\bibinfo {volume} {120}},\
  \bibinfo {pages} {106601} (\bibinfo {year} {2018})}\BibitemShut {NoStop}%
\bibitem [{\citenamefont {Kolodrubetz}\ \emph {et~al.}(2018)\citenamefont
  {Kolodrubetz}, \citenamefont {Nathan}, \citenamefont {Gazit}, \citenamefont
  {Morimoto},\ and\ \citenamefont {Moore}}]{Moore_PRL18}%
  \BibitemOpen
  \bibfield  {author} {\bibinfo {author} {\bibfnamefont {M.~H.}\ \bibnamefont
  {Kolodrubetz}}, \bibinfo {author} {\bibfnamefont {F.}~\bibnamefont {Nathan}},
  \bibinfo {author} {\bibfnamefont {S.}~\bibnamefont {Gazit}}, \bibinfo
  {author} {\bibfnamefont {T.}~\bibnamefont {Morimoto}}, \ and\ \bibinfo
  {author} {\bibfnamefont {J.~E.}\ \bibnamefont {Moore}},\ }\href {\doibase
  10.1103/PhysRevLett.120.150601} {\bibfield  {journal} {\bibinfo  {journal}
  {Phys. Rev. Lett.}\ }\textbf {\bibinfo {volume} {120}},\ \bibinfo {pages}
  {150601} (\bibinfo {year} {2018})}\BibitemShut {NoStop}%
\bibitem [{\citenamefont {Nakagawa}\ \emph {et~al.}(2018)\citenamefont
  {Nakagawa}, \citenamefont {Yoshida}, \citenamefont {Peters},\ and\
  \citenamefont {Kawakami}}]{Nakagawa_PRB18}%
  \BibitemOpen
  \bibfield  {author} {\bibinfo {author} {\bibfnamefont {M.}~\bibnamefont
  {Nakagawa}}, \bibinfo {author} {\bibfnamefont {T.}~\bibnamefont {Yoshida}},
  \bibinfo {author} {\bibfnamefont {R.}~\bibnamefont {Peters}}, \ and\ \bibinfo
  {author} {\bibfnamefont {N.}~\bibnamefont {Kawakami}},\ }\href {\doibase
  10.1103/PhysRevB.98.115147} {\bibfield  {journal} {\bibinfo  {journal} {Phys.
  Rev. B}\ }\textbf {\bibinfo {volume} {98}},\ \bibinfo {pages} {115147}
  (\bibinfo {year} {2018})}\BibitemShut {NoStop}%
\bibitem [{\citenamefont {Haug}\ \emph {et~al.}(2019)\citenamefont {Haug},
  \citenamefont {Amico}, \citenamefont {Kwek}, \citenamefont {Munro},\ and\
  \citenamefont {Bastidas}}]{Haug_arxiv19}%
  \BibitemOpen
  \bibfield  {author} {\bibinfo {author} {\bibfnamefont {T.}~\bibnamefont
  {Haug}}, \bibinfo {author} {\bibfnamefont {L.}~\bibnamefont {Amico}},
  \bibinfo {author} {\bibfnamefont {L.-C.}\ \bibnamefont {Kwek}}, \bibinfo
  {author} {\bibfnamefont {W.}~\bibnamefont {Munro}}, \ and\ \bibinfo {author}
  {\bibfnamefont {V.}~\bibnamefont {Bastidas}},\ }\href@noop {} {\bibfield
  {journal} {\bibinfo  {journal} {arXiv preprint arXiv:1905.03807}\ } (\bibinfo
  {year} {2019})}\BibitemShut {NoStop}%
\bibitem [{\citenamefont {Arceci}\ \emph {et~al.}(2019)\citenamefont {Arceci},
  \citenamefont {Russomanno},\ and\ \citenamefont {Santoro}}]{Arceci_Arx19}%
  \BibitemOpen
  \bibfield  {author} {\bibinfo {author} {\bibfnamefont {L.}~\bibnamefont
  {Arceci}}, \bibinfo {author} {\bibfnamefont {A.}~\bibnamefont {Russomanno}},
  \ and\ \bibinfo {author} {\bibfnamefont {G.~E.}\ \bibnamefont {Santoro}},\
  }\href {https://arxiv.org/abs/1905.08808} {\bibfield  {journal} {\bibinfo
  {journal} {arXiv preprint}\ ,\ \bibinfo {eid} {arXiv:1905.08808}} (\bibinfo
  {year} {2019})}\BibitemShut {NoStop}%
\bibitem [{\citenamefont {Lu}\ \emph {et~al.}(2014)\citenamefont {Lu},
  \citenamefont {Joannopoulos},\ and\ \citenamefont
  {Soljačić}}]{Lu_NatPhot14}%
  \BibitemOpen
  \bibfield  {author} {\bibinfo {author} {\bibfnamefont {L.}~\bibnamefont
  {Lu}}, \bibinfo {author} {\bibfnamefont {J.~D.}\ \bibnamefont
  {Joannopoulos}}, \ and\ \bibinfo {author} {\bibfnamefont {M.}~\bibnamefont
  {Soljačić}},\ }\href {\doibase 10.1038/nphoton.2014.248} {\bibfield
  {journal} {\bibinfo  {journal} {Nature Photonics}\ }\textbf {\bibinfo
  {volume} {8}},\ \bibinfo {pages} {821} (\bibinfo {year} {2014})}\BibitemShut
  {NoStop}%
\bibitem [{\citenamefont {Lohse}\ \emph {et~al.}(2016)\citenamefont {Lohse},
  \citenamefont {Schweizer}, \citenamefont {Zilberberg}, \citenamefont
  {Aidelsburger},\ and\ \citenamefont {Bloch}}]{Lohse_Nphys16}%
  \BibitemOpen
  \bibfield  {author} {\bibinfo {author} {\bibfnamefont {M.}~\bibnamefont
  {Lohse}}, \bibinfo {author} {\bibfnamefont {C.}~\bibnamefont {Schweizer}},
  \bibinfo {author} {\bibfnamefont {O.}~\bibnamefont {Zilberberg}}, \bibinfo
  {author} {\bibfnamefont {M.}~\bibnamefont {Aidelsburger}}, \ and\ \bibinfo
  {author} {\bibfnamefont {I.}~\bibnamefont {Bloch}},\ }\href@noop {}
  {\bibfield  {journal} {\bibinfo  {journal} {Nat. Phys.}\ }\textbf {\bibinfo
  {volume} {12}},\ \bibinfo {pages} {350} (\bibinfo {year} {2016})}\BibitemShut
  {NoStop}%
\bibitem [{\citenamefont {Nakajima}\ \emph {et~al.}(2016)\citenamefont
  {Nakajima}, \citenamefont {Tomita}, \citenamefont {Taie}, \citenamefont
  {Ichinose}, \citenamefont {Ozawa}, \citenamefont {Wang}, \citenamefont
  {Troyer},\ and\ \citenamefont {Takahashi}}]{Nakajima_Nphys16}%
  \BibitemOpen
  \bibfield  {author} {\bibinfo {author} {\bibfnamefont {S.}~\bibnamefont
  {Nakajima}}, \bibinfo {author} {\bibfnamefont {T.}~\bibnamefont {Tomita}},
  \bibinfo {author} {\bibfnamefont {S.}~\bibnamefont {Taie}}, \bibinfo {author}
  {\bibfnamefont {T.}~\bibnamefont {Ichinose}}, \bibinfo {author}
  {\bibfnamefont {H.}~\bibnamefont {Ozawa}}, \bibinfo {author} {\bibfnamefont
  {L.}~\bibnamefont {Wang}}, \bibinfo {author} {\bibfnamefont {M.}~\bibnamefont
  {Troyer}}, \ and\ \bibinfo {author} {\bibfnamefont {Y.}~\bibnamefont
  {Takahashi}},\ }\href@noop {} {\bibfield  {journal} {\bibinfo  {journal}
  {Nat. Phys.}\ }\textbf {\bibinfo {volume} {12}},\ \bibinfo {pages} {296}
  (\bibinfo {year} {2016})}\BibitemShut {NoStop}%
\bibitem [{\citenamefont {Lohse}\ \emph {et~al.}(2018)\citenamefont {Lohse},
  \citenamefont {Schweizer}, \citenamefont {Price}, \citenamefont
  {Zilberberg},\ and\ \citenamefont {Bloch}}]{Lohse_Nat18}%
  \BibitemOpen
  \bibfield  {author} {\bibinfo {author} {\bibfnamefont {M.}~\bibnamefont
  {Lohse}}, \bibinfo {author} {\bibfnamefont {C.}~\bibnamefont {Schweizer}},
  \bibinfo {author} {\bibfnamefont {H.~M.}\ \bibnamefont {Price}}, \bibinfo
  {author} {\bibfnamefont {O.}~\bibnamefont {Zilberberg}}, \ and\ \bibinfo
  {author} {\bibfnamefont {I.}~\bibnamefont {Bloch}},\ }\href {\doibase
  10.1038/nature25000} {\bibfield  {journal} {\bibinfo  {journal} {Nature}\
  }\textbf {\bibinfo {volume} {553}},\ \bibinfo {pages} {55} (\bibinfo {year}
  {2018})}\BibitemShut {NoStop}%
\bibitem [{\citenamefont {Ma}\ \emph {et~al.}(2018)\citenamefont {Ma},
  \citenamefont {Zhou}, \citenamefont {Zhang}, \citenamefont {Li},
  \citenamefont {Cheng}, \citenamefont {Geng}, \citenamefont {Rong},
  \citenamefont {Shi}, \citenamefont {Gong},\ and\ \citenamefont
  {Du}}]{Ma_PRL18}%
  \BibitemOpen
  \bibfield  {author} {\bibinfo {author} {\bibfnamefont {W.}~\bibnamefont
  {Ma}}, \bibinfo {author} {\bibfnamefont {L.}~\bibnamefont {Zhou}}, \bibinfo
  {author} {\bibfnamefont {Q.}~\bibnamefont {Zhang}}, \bibinfo {author}
  {\bibfnamefont {M.}~\bibnamefont {Li}}, \bibinfo {author} {\bibfnamefont
  {C.}~\bibnamefont {Cheng}}, \bibinfo {author} {\bibfnamefont
  {J.}~\bibnamefont {Geng}}, \bibinfo {author} {\bibfnamefont {X.}~\bibnamefont
  {Rong}}, \bibinfo {author} {\bibfnamefont {F.}~\bibnamefont {Shi}}, \bibinfo
  {author} {\bibfnamefont {J.}~\bibnamefont {Gong}}, \ and\ \bibinfo {author}
  {\bibfnamefont {J.}~\bibnamefont {Du}},\ }\href {\doibase
  10.1103/PhysRevLett.120.120501} {\bibfield  {journal} {\bibinfo  {journal}
  {Phys. Rev. Lett.}\ }\textbf {\bibinfo {volume} {120}},\ \bibinfo {pages}
  {120501} (\bibinfo {year} {2018})}\BibitemShut {NoStop}%
\bibitem [{\citenamefont {{Hotzen Grinberg}}\ \emph {et~al.}(2019)\citenamefont
  {{Hotzen Grinberg}}, \citenamefont {{Lin}}, \citenamefont {{Harris}},
  \citenamefont {{Benalcazar}}, \citenamefont {{Peterson}}, \citenamefont
  {{Hughes}},\ and\ \citenamefont {{Bahl}}}]{Hotzen_Arx19}%
  \BibitemOpen
  \bibfield  {author} {\bibinfo {author} {\bibfnamefont {I.}~\bibnamefont
  {{Hotzen Grinberg}}}, \bibinfo {author} {\bibfnamefont {M.}~\bibnamefont
  {{Lin}}}, \bibinfo {author} {\bibfnamefont {C.}~\bibnamefont {{Harris}}},
  \bibinfo {author} {\bibfnamefont {W.~A.}\ \bibnamefont {{Benalcazar}}},
  \bibinfo {author} {\bibfnamefont {C.~W.}\ \bibnamefont {{Peterson}}},
  \bibinfo {author} {\bibfnamefont {T.~L.}\ \bibnamefont {{Hughes}}}, \ and\
  \bibinfo {author} {\bibfnamefont {G.}~\bibnamefont {{Bahl}}},\ }\href@noop {}
  {\bibfield  {journal} {\bibinfo  {journal} {arXiv e-prints}\ ,\ \bibinfo
  {pages} {arXiv:1905.02778}} (\bibinfo {year} {2019})},\ \Eprint
  {http://arxiv.org/abs/1905.02778} {1905.02778} \BibitemShut {NoStop}%
\bibitem [{\citenamefont {Rudner}\ \emph {et~al.}(2013)\citenamefont {Rudner},
  \citenamefont {Lindner}, \citenamefont {Berg},\ and\ \citenamefont
  {Levin}}]{rudner2013anomalous}%
  \BibitemOpen
  \bibfield  {author} {\bibinfo {author} {\bibfnamefont {M.~S.}\ \bibnamefont
  {Rudner}}, \bibinfo {author} {\bibfnamefont {N.~H.}\ \bibnamefont {Lindner}},
  \bibinfo {author} {\bibfnamefont {E.}~\bibnamefont {Berg}}, \ and\ \bibinfo
  {author} {\bibfnamefont {M.}~\bibnamefont {Levin}},\ }\href@noop {}
  {\bibfield  {journal} {\bibinfo  {journal} {Phys. Rev. X}\ }\textbf {\bibinfo
  {volume} {3}},\ \bibinfo {pages} {031005} (\bibinfo {year}
  {2013})}\BibitemShut {NoStop}%
\bibitem [{\citenamefont {Potter}\ \emph {et~al.}(2016)\citenamefont {Potter},
  \citenamefont {Morimoto},\ and\ \citenamefont {Vishwanath}}]{Potter_PRX16}%
  \BibitemOpen
  \bibfield  {author} {\bibinfo {author} {\bibfnamefont {A.~C.}\ \bibnamefont
  {Potter}}, \bibinfo {author} {\bibfnamefont {T.}~\bibnamefont {Morimoto}}, \
  and\ \bibinfo {author} {\bibfnamefont {A.}~\bibnamefont {Vishwanath}},\
  }\href@noop {} {\bibfield  {journal} {\bibinfo  {journal} {Physical Review
  X}\ }\textbf {\bibinfo {volume} {6}},\ \bibinfo {pages} {041001} (\bibinfo
  {year} {2016})}\BibitemShut {NoStop}%
\bibitem [{\citenamefont {Else}\ and\ \citenamefont
  {Nayak}(2016)}]{Else_PRB16}%
  \BibitemOpen
  \bibfield  {author} {\bibinfo {author} {\bibfnamefont {D.~V.}\ \bibnamefont
  {Else}}\ and\ \bibinfo {author} {\bibfnamefont {C.}~\bibnamefont {Nayak}},\
  }\href@noop {} {\bibfield  {journal} {\bibinfo  {journal} {Physical Review
  B}\ }\textbf {\bibinfo {volume} {93}},\ \bibinfo {pages} {201103} (\bibinfo
  {year} {2016})}\BibitemShut {NoStop}%
\bibitem [{\citenamefont {Budich}\ \emph {et~al.}(2017)\citenamefont {Budich},
  \citenamefont {Hu},\ and\ \citenamefont {Zoller}}]{Budich_PRL17}%
  \BibitemOpen
  \bibfield  {author} {\bibinfo {author} {\bibfnamefont {J.~C.}\ \bibnamefont
  {Budich}}, \bibinfo {author} {\bibfnamefont {Y.}~\bibnamefont {Hu}}, \ and\
  \bibinfo {author} {\bibfnamefont {P.}~\bibnamefont {Zoller}},\ }\href@noop {}
  {\bibfield  {journal} {\bibinfo  {journal} {Phys. Rev. Lett.}\ }\textbf
  {\bibinfo {volume} {118}},\ \bibinfo {pages} {105302} (\bibinfo {year}
  {2017})}\BibitemShut {NoStop}%
\bibitem [{\citenamefont {Esin}\ \emph {et~al.}(2018)\citenamefont {Esin},
  \citenamefont {Rudner}, \citenamefont {Refael},\ and\ \citenamefont
  {Lindner}}]{Lindner_PRB18}%
  \BibitemOpen
  \bibfield  {author} {\bibinfo {author} {\bibfnamefont {I.}~\bibnamefont
  {Esin}}, \bibinfo {author} {\bibfnamefont {M.~S.}\ \bibnamefont {Rudner}},
  \bibinfo {author} {\bibfnamefont {G.}~\bibnamefont {Refael}}, \ and\ \bibinfo
  {author} {\bibfnamefont {N.~H.}\ \bibnamefont {Lindner}},\ }\href {\doibase
  10.1103/PhysRevB.97.245401} {\bibfield  {journal} {\bibinfo  {journal} {Phys.
  Rev. B}\ }\textbf {\bibinfo {volume} {97}},\ \bibinfo {pages} {245401}
  (\bibinfo {year} {2018})}\BibitemShut {NoStop}%
\bibitem [{\citenamefont {Shtanko}\ and\ \citenamefont
  {Movassagh}(2018)}]{Shtanko_PRL18}%
  \BibitemOpen
  \bibfield  {author} {\bibinfo {author} {\bibfnamefont {O.}~\bibnamefont
  {Shtanko}}\ and\ \bibinfo {author} {\bibfnamefont {R.}~\bibnamefont
  {Movassagh}},\ }\href {\doibase 10.1103/PhysRevLett.121.126803} {\bibfield
  {journal} {\bibinfo  {journal} {Phys. Rev. Lett.}\ }\textbf {\bibinfo
  {volume} {121}},\ \bibinfo {pages} {126803} (\bibinfo {year}
  {2018})}\BibitemShut {NoStop}%
\bibitem [{\citenamefont {Lindner}\ \emph {et~al.}(2017)\citenamefont
  {Lindner}, \citenamefont {Berg},\ and\ \citenamefont
  {Rudner}}]{Lindner_PRX17}%
  \BibitemOpen
  \bibfield  {author} {\bibinfo {author} {\bibfnamefont {N.~H.}\ \bibnamefont
  {Lindner}}, \bibinfo {author} {\bibfnamefont {E.}~\bibnamefont {Berg}}, \
  and\ \bibinfo {author} {\bibfnamefont {M.~S.}\ \bibnamefont {Rudner}},\
  }\href@noop {} {\bibfield  {journal} {\bibinfo  {journal} {Phys. Rev. X}\
  }\textbf {\bibinfo {volume} {7}},\ \bibinfo {pages} {011018} (\bibinfo {year}
  {2017})}\BibitemShut {NoStop}%
\bibitem [{\citenamefont {Gulden}\ \emph {et~al.}(2019)\citenamefont {Gulden},
  \citenamefont {Berg}, \citenamefont {Rudner},\ and\ \citenamefont
  {Lindner}}]{Gulden_arxiv19}%
  \BibitemOpen
  \bibfield  {author} {\bibinfo {author} {\bibfnamefont {T.}~\bibnamefont
  {Gulden}}, \bibinfo {author} {\bibfnamefont {E.}~\bibnamefont {Berg}},
  \bibinfo {author} {\bibfnamefont {M.~S.}\ \bibnamefont {Rudner}}, \ and\
  \bibinfo {author} {\bibfnamefont {N.~H.}\ \bibnamefont {Lindner}},\
  }\href@noop {} {\bibfield  {journal} {\bibinfo  {journal} {arXiv preprint
  arXiv:1901.08385}\ } (\bibinfo {year} {2019})}\BibitemShut {NoStop}%
\bibitem [{\citenamefont {Abanin}\ \emph {et~al.}(2018)\citenamefont {Abanin},
  \citenamefont {Altman}, \citenamefont {Bloch},\ and\ \citenamefont
  {Serbyn}}]{Bloch1}%
  \BibitemOpen
  \bibfield  {author} {\bibinfo {author} {\bibfnamefont {D.~A.}\ \bibnamefont
  {Abanin}}, \bibinfo {author} {\bibfnamefont {E.}~\bibnamefont {Altman}},
  \bibinfo {author} {\bibfnamefont {I.}~\bibnamefont {Bloch}}, \ and\ \bibinfo
  {author} {\bibfnamefont {M.}~\bibnamefont {Serbyn}},\ }\href
  {https://arxiv.org/abs/1804.11065} {\bibfield  {journal} {\bibinfo  {journal}
  {arXiv}\ ,\ \bibinfo {pages} {1804.11065}} (\bibinfo {year}
  {2018})}\BibitemShut {NoStop}%
\bibitem [{\citenamefont {Harper}\ \emph {et~al.}(2019)\citenamefont {Harper},
  \citenamefont {Roy}, \citenamefont {Rudner},\ and\ \citenamefont
  {Sondhi}}]{Harper_arxiv19}%
  \BibitemOpen
  \bibfield  {author} {\bibinfo {author} {\bibfnamefont {F.}~\bibnamefont
  {Harper}}, \bibinfo {author} {\bibfnamefont {R.}~\bibnamefont {Roy}},
  \bibinfo {author} {\bibfnamefont {M.~S.}\ \bibnamefont {Rudner}}, \ and\
  \bibinfo {author} {\bibfnamefont {S.}~\bibnamefont {Sondhi}},\ }\href@noop {}
  {\bibfield  {journal} {\bibinfo  {journal} {arXiv preprint arXiv:1905.01317}\
  } (\bibinfo {year} {2019})}\BibitemShut {NoStop}%
\bibitem [{\citenamefont {Hatami}\ \emph {et~al.}(2016)\citenamefont {Hatami},
  \citenamefont {Danieli}, \citenamefont {Bodyfelt},\ and\ \citenamefont
  {Flach}}]{Hatami_PRE16}%
  \BibitemOpen
  \bibfield  {author} {\bibinfo {author} {\bibfnamefont {H.}~\bibnamefont
  {Hatami}}, \bibinfo {author} {\bibfnamefont {C.}~\bibnamefont {Danieli}},
  \bibinfo {author} {\bibfnamefont {J.~D.}\ \bibnamefont {Bodyfelt}}, \ and\
  \bibinfo {author} {\bibfnamefont {S.}~\bibnamefont {Flach}},\ }\href@noop {}
  {\bibfield  {journal} {\bibinfo  {journal} {Physical Review E}\ }\textbf
  {\bibinfo {volume} {93}} (\bibinfo {year} {2016})}\BibitemShut {NoStop}%
\bibitem [{\citenamefont {Agarwal}\ \emph {et~al.}(2017)\citenamefont
  {Agarwal}, \citenamefont {Ganeshan},\ and\ \citenamefont
  {Bhatt}}]{Agarwal_PRB17}%
  \BibitemOpen
  \bibfield  {author} {\bibinfo {author} {\bibfnamefont {K.}~\bibnamefont
  {Agarwal}}, \bibinfo {author} {\bibfnamefont {S.}~\bibnamefont {Ganeshan}}, \
  and\ \bibinfo {author} {\bibfnamefont {R.~N.}\ \bibnamefont {Bhatt}},\ }\href
  {\doibase 10.1103/PhysRevB.96.014201} {\bibfield  {journal} {\bibinfo
  {journal} {Phys. Rev. B}\ }\textbf {\bibinfo {volume} {96}},\ \bibinfo
  {pages} {014201} (\bibinfo {year} {2017})}\BibitemShut {NoStop}%
\bibitem [{\citenamefont {Rice}\ and\ \citenamefont {Mele}(1982)}]{Rice_PRL82}%
  \BibitemOpen
  \bibfield  {author} {\bibinfo {author} {\bibfnamefont {M.}~\bibnamefont
  {Rice}}\ and\ \bibinfo {author} {\bibfnamefont {E.}~\bibnamefont {Mele}},\
  }\href@noop {} {\bibfield  {journal} {\bibinfo  {journal} {Phys. Rev. Lett.}\
  }\textbf {\bibinfo {volume} {49}},\ \bibinfo {pages} {1455} (\bibinfo {year}
  {1982})}\BibitemShut {NoStop}%
\bibitem [{\citenamefont {Avron}\ and\ \citenamefont
  {Kons}(1999)}]{Avron_JPA99}%
  \BibitemOpen
  \bibfield  {author} {\bibinfo {author} {\bibfnamefont {J.~E.}\ \bibnamefont
  {Avron}}\ and\ \bibinfo {author} {\bibfnamefont {Z.}~\bibnamefont {Kons}},\
  }\href@noop {} {\bibfield  {journal} {\bibinfo  {journal} {J. Phys. A-Math.
  Gen.}\ }\textbf {\bibinfo {volume} {32}},\ \bibinfo {pages} {6097} (\bibinfo
  {year} {1999})}\BibitemShut {NoStop}%
\bibitem [{\citenamefont {Grifoni}\ and\ \citenamefont
  {H{\"a}nggi}(1998)}]{Grifoni1998driven}%
  \BibitemOpen
  \bibfield  {author} {\bibinfo {author} {\bibfnamefont {M.}~\bibnamefont
  {Grifoni}}\ and\ \bibinfo {author} {\bibfnamefont {P.}~\bibnamefont
  {H{\"a}nggi}},\ }\href@noop {} {\bibfield  {journal} {\bibinfo  {journal}
  {Physics Reports}\ }\textbf {\bibinfo {volume} {304}},\ \bibinfo {pages}
  {229} (\bibinfo {year} {1998})}\BibitemShut {NoStop}%
\bibitem [{\citenamefont {Russomanno}\ \emph {et~al.}(2012)\citenamefont
  {Russomanno}, \citenamefont {Silva},\ and\ \citenamefont
  {Santoro}}]{Russomanno_PRL12}%
  \BibitemOpen
  \bibfield  {author} {\bibinfo {author} {\bibfnamefont {A.}~\bibnamefont
  {Russomanno}}, \bibinfo {author} {\bibfnamefont {A.}~\bibnamefont {Silva}}, \
  and\ \bibinfo {author} {\bibfnamefont {G.~E.}\ \bibnamefont {Santoro}},\
  }\href@noop {} {\bibfield  {journal} {\bibinfo  {journal} {Phys. Rev. Lett.}\
  }\textbf {\bibinfo {volume} {109}},\ \bibinfo {pages} {257201} (\bibinfo
  {year} {2012})}\BibitemShut {NoStop}%
\bibitem [{\citenamefont {Abrahams}\ \emph {et~al.}(1979)\citenamefont
  {Abrahams}, \citenamefont {Anderson}, \citenamefont {Licciardello},\ and\
  \citenamefont {Ramakrishnan}}]{Abrahams_PRL79}%
  \BibitemOpen
  \bibfield  {author} {\bibinfo {author} {\bibfnamefont {E.}~\bibnamefont
  {Abrahams}}, \bibinfo {author} {\bibfnamefont {P.~W.}\ \bibnamefont
  {Anderson}}, \bibinfo {author} {\bibfnamefont {D.~C.}\ \bibnamefont
  {Licciardello}}, \ and\ \bibinfo {author} {\bibfnamefont {T.~V.}\
  \bibnamefont {Ramakrishnan}},\ }\href {\doibase 10.1103/PhysRevLett.42.673}
  {\bibfield  {journal} {\bibinfo  {journal} {Phys. Rev. Lett.}\ }\textbf
  {\bibinfo {volume} {42}},\ \bibinfo {pages} {673} (\bibinfo {year}
  {1979})}\BibitemShut {NoStop}%
\bibitem [{\citenamefont {Edwards}\ and\ \citenamefont
  {Thouless}(1972)}]{Edwards_JPC72}%
  \BibitemOpen
  \bibfield  {author} {\bibinfo {author} {\bibfnamefont {J.~T.}\ \bibnamefont
  {Edwards}}\ and\ \bibinfo {author} {\bibfnamefont {D.~J.}\ \bibnamefont
  {Thouless}},\ }\href@noop {} {\bibfield  {journal} {\bibinfo  {journal} {J.
  Phys. C}\ }\textbf {\bibinfo {volume} {5}},\ \bibinfo {pages} {807} (\bibinfo
  {year} {1972})}\BibitemShut {NoStop}%
\bibitem [{nota1()}]{nota1}%
  \BibitemOpen
  \bibinfo {note} {Incidentally, the average IPR suggests a localized
  scenario, as $\overline{\IPR}=\frac{1}{L} \sum_\alpha \IPR_\alpha$ saturates
  to a finite value for $L\to \infty$}\BibitemShut {NoStop}%
\bibitem [{\citenamefont {Citro}\ \emph {et~al.}()\citenamefont {Citro},
  \citenamefont {Wauters}, \citenamefont {Privitera}, \citenamefont
  {Russomanno},\ and\ \citenamefont {Santoro}}]{Citro_unpub}%
  \BibitemOpen
  \bibfield  {author} {\bibinfo {author} {\bibfnamefont {R.}~\bibnamefont
  {Citro}}, \bibinfo {author} {\bibfnamefont {M.~M.}\ \bibnamefont {Wauters}},
  \bibinfo {author} {\bibfnamefont {L.}~\bibnamefont {Privitera}}, \bibinfo
  {author} {\bibfnamefont {A.}~\bibnamefont {Russomanno}}, \ and\ \bibinfo
  {author} {\bibfnamefont {G.~E.}\ \bibnamefont {Santoro}},\ }\href@noop {}
  {}\bibinfo {note} {(in preparation)}\BibitemShut {NoStop}%
\bibitem [{\citenamefont {Richard E.~Prange}(1990)}]{Prange_book90}%
  \BibitemOpen
  \bibfield  {author} {\bibinfo {author} {\bibfnamefont {S.~M.~G.}\
  \bibnamefont {Richard E.~Prange}},\ }\href@noop {} {\emph {\bibinfo {title}
  {The Quantum Hall Effect}}},\ \bibinfo {edition} {2nd}\ ed.,\ Graduate Texts
  in Contemporary Physics\ (\bibinfo  {publisher} {Springer-Verlag New York},\
  \bibinfo {year} {1990})\BibitemShut {NoStop}%
\bibitem [{\citenamefont {Prodan}\ \emph {et~al.}(2010)\citenamefont {Prodan},
  \citenamefont {Hughes},\ and\ \citenamefont {Bernevig}}]{Prodan_PRL2010}%
  \BibitemOpen
  \bibfield  {author} {\bibinfo {author} {\bibfnamefont {E.}~\bibnamefont
  {Prodan}}, \bibinfo {author} {\bibfnamefont {T.~L.}\ \bibnamefont {Hughes}},
  \ and\ \bibinfo {author} {\bibfnamefont {B.~A.}\ \bibnamefont {Bernevig}},\
  }\href {\doibase 10.1103/PhysRevLett.105.115501} {\bibfield  {journal}
  {\bibinfo  {journal} {Phys. Rev. Lett.}\ }\textbf {\bibinfo {volume} {105}},\
  \bibinfo {pages} {115501} (\bibinfo {year} {2010})}\BibitemShut {NoStop}%
\bibitem [{\citenamefont {Prodan}(2011)}]{Prodan_JPA2011}%
  \BibitemOpen
  \bibfield  {author} {\bibinfo {author} {\bibfnamefont {E.}~\bibnamefont
  {Prodan}},\ }\href {\doibase 10.1088/1751-8113/44/11/113001} {\bibfield
  {journal} {\bibinfo  {journal} {Journal of Physics A: Mathematical and
  Theoretical}\ }\textbf {\bibinfo {volume} {44}},\ \bibinfo {pages} {113001}
  (\bibinfo {year} {2011})}\BibitemShut {NoStop}%
\bibitem [{\citenamefont {Yang}\ and\ \citenamefont
  {Bhatt}(1996)}]{Yang_PRL96}%
  \BibitemOpen
  \bibfield  {author} {\bibinfo {author} {\bibfnamefont {K.}~\bibnamefont
  {Yang}}\ and\ \bibinfo {author} {\bibfnamefont {R.~N.}\ \bibnamefont
  {Bhatt}},\ }\href {\doibase 10.1103/PhysRevLett.76.1316} {\bibfield
  {journal} {\bibinfo  {journal} {Phys. Rev. Lett.}\ }\textbf {\bibinfo
  {volume} {76}},\ \bibinfo {pages} {1316} (\bibinfo {year}
  {1996})}\BibitemShut {NoStop}%
\bibitem [{\citenamefont {Avron}\ \emph {et~al.}(2003)\citenamefont {Avron},
  \citenamefont {Osadchy},\ and\ \citenamefont {Seiler}}]{Avron_PT03}%
  \BibitemOpen
  \bibfield  {author} {\bibinfo {author} {\bibfnamefont {J.~E.}\ \bibnamefont
  {Avron}}, \bibinfo {author} {\bibfnamefont {D.}~\bibnamefont {Osadchy}}, \
  and\ \bibinfo {author} {\bibfnamefont {R.}~\bibnamefont {Seiler}},\
  }\href@noop {} {\bibfield  {journal} {\bibinfo  {journal} {Phys. Today}\
  }\textbf {\bibinfo {volume} {56}},\ \bibinfo {pages} {38} (\bibinfo {year}
  {2003})}\BibitemShut {NoStop}%
\bibitem [{\citenamefont {Martin}\ \emph {et~al.}(2017)\citenamefont {Martin},
  \citenamefont {Refael},\ and\ \citenamefont {Halperin}}]{Martin_PRX17}%
  \BibitemOpen
  \bibfield  {author} {\bibinfo {author} {\bibfnamefont {I.}~\bibnamefont
  {Martin}}, \bibinfo {author} {\bibfnamefont {G.}~\bibnamefont {Refael}}, \
  and\ \bibinfo {author} {\bibfnamefont {B.}~\bibnamefont {Halperin}},\ }\href
  {\doibase 10.1103/PhysRevX.7.041008} {\bibfield  {journal} {\bibinfo
  {journal} {Phys. Rev. X}\ }\textbf {\bibinfo {volume} {7}},\ \bibinfo {pages}
  {041008} (\bibinfo {year} {2017})}\BibitemShut {NoStop}%
\bibitem [{\citenamefont {Shih}\ and\ \citenamefont {Niu}(1994)}]{Shih_PRB94}%
  \BibitemOpen
  \bibfield  {author} {\bibinfo {author} {\bibfnamefont {W.-K.}\ \bibnamefont
  {Shih}}\ and\ \bibinfo {author} {\bibfnamefont {Q.}~\bibnamefont {Niu}},\
  }\href@noop {} {\bibfield  {journal} {\bibinfo  {journal} {Phys. Rev. B}\
  }\textbf {\bibinfo {volume} {50}},\ \bibinfo {pages} {11902} (\bibinfo {year}
  {1994})}\BibitemShut {NoStop}%
\bibitem [{\citenamefont {Ferrari}(1998)}]{Ferrari_IJMP98}%
  \BibitemOpen
  \bibfield  {author} {\bibinfo {author} {\bibfnamefont {R.}~\bibnamefont
  {Ferrari}},\ }\href@noop {} {\bibfield  {journal} {\bibinfo  {journal} {Int.
  J. Mod. Phys. B}\ }\textbf {\bibinfo {volume} {12}},\ \bibinfo {pages} {1105}
  (\bibinfo {year} {1998})}\BibitemShut {NoStop}%
\bibitem [{\citenamefont {Watanabe}(2018)}]{Watanabe_PRB18}%
  \BibitemOpen
  \bibfield  {author} {\bibinfo {author} {\bibfnamefont {H.}~\bibnamefont
  {Watanabe}},\ }\href {\doibase 10.1103/PhysRevB.98.155137} {\bibfield
  {journal} {\bibinfo  {journal} {Phys. Rev. B}\ }\textbf {\bibinfo {volume}
  {98}},\ \bibinfo {pages} {155137} (\bibinfo {year} {2018})}\BibitemShut
  {NoStop}%
\bibitem [{\citenamefont {Wauters}\ and\ \citenamefont
  {Santoro}(2018)}]{Wauters_PRB18}%
  \BibitemOpen
  \bibfield  {author} {\bibinfo {author} {\bibfnamefont {M.~M.}\ \bibnamefont
  {Wauters}}\ and\ \bibinfo {author} {\bibfnamefont {G.~E.}\ \bibnamefont
  {Santoro}},\ }\href {\doibase 10.1103/PhysRevB.98.205112} {\bibfield
  {journal} {\bibinfo  {journal} {Phys. Rev. B}\ }\textbf {\bibinfo {volume}
  {98}},\ \bibinfo {pages} {205112} (\bibinfo {year} {2018})}\BibitemShut
  {NoStop}%
\bibitem [{nota2()}]{nota2}%
  \BibitemOpen
  \bibinfo {note} {If disorder induces true level crossings in the occupied
  band of the SP spectrum, as in the {\em control freak
  limit}~\cite{Asboth_book15}, a fine tuned driving is able to mix localized
  energy eigenstates into extended Floquet states, even at vanishing
  frequency.}\BibitemShut {Stop}%
\bibitem [{\citenamefont {Khemani}\ \emph {et~al.}(2015)\citenamefont
  {Khemani}, \citenamefont {Nandkishore},\ and\ \citenamefont
  {Sondhi}}]{Khemani_NatPhys15}%
  \BibitemOpen
  \bibfield  {author} {\bibinfo {author} {\bibfnamefont {V.}~\bibnamefont
  {Khemani}}, \bibinfo {author} {\bibfnamefont {R.}~\bibnamefont
  {Nandkishore}}, \ and\ \bibinfo {author} {\bibfnamefont {S.~L.}\ \bibnamefont
  {Sondhi}},\ }\href {\doibase 10.1038/nphys3344} {\bibfield  {journal}
  {\bibinfo  {journal} {Nature Physics}\ }\textbf {\bibinfo {volume} {11}},\
  \bibinfo {pages} {560} (\bibinfo {year} {2015})}\BibitemShut {NoStop}%
\bibitem [{\citenamefont {Ippoliti}\ and\ \citenamefont
  {Bhatt}(2019)}]{Ippoliti_arx19}%
  \BibitemOpen
  \bibfield  {author} {\bibinfo {author} {\bibfnamefont {M.}~\bibnamefont
  {Ippoliti}}\ and\ \bibinfo {author} {\bibfnamefont {R.~N.}\ \bibnamefont
  {Bhatt}},\ }\href {https://arxiv.org/abs/1905.13171} {\bibfield  {journal}
  {\bibinfo  {journal} {arXiv:1905.13171}\ } (\bibinfo {year}
  {2019})}\BibitemShut {NoStop}%
\bibitem [{\citenamefont {Bordia}\ \emph {et~al.}(2017)\citenamefont {Bordia},
  \citenamefont {Lüschen}, \citenamefont {Schneider}, \citenamefont {Knap},\
  and\ \citenamefont {Bloch}}]{Bordia_NatPhys17}%
  \BibitemOpen
  \bibfield  {author} {\bibinfo {author} {\bibfnamefont {P.}~\bibnamefont
  {Bordia}}, \bibinfo {author} {\bibfnamefont {H.}~\bibnamefont {Lüschen}},
  \bibinfo {author} {\bibfnamefont {U.}~\bibnamefont {Schneider}}, \bibinfo
  {author} {\bibfnamefont {M.}~\bibnamefont {Knap}}, \ and\ \bibinfo {author}
  {\bibfnamefont {I.}~\bibnamefont {Bloch}},\ }\href {\doibase
  10.1038/nphys4020} {\bibfield  {journal} {\bibinfo  {journal} {Nature
  Physics}\ }\textbf {\bibinfo {volume} {13}},\ \bibinfo {pages} {460}
  (\bibinfo {year} {2017})}\BibitemShut {NoStop}%
\bibitem [{\citenamefont {Asboth}\ \emph {et~al.}(2016)\citenamefont {Asboth},
  \citenamefont {Oroszlany},\ and\ \citenamefont {Palyi}}]{Asboth_book15}%
  \BibitemOpen
  \bibfield  {author} {\bibinfo {author} {\bibfnamefont {J.~K.}\ \bibnamefont
  {Asboth}}, \bibinfo {author} {\bibfnamefont {L.}~\bibnamefont {Oroszlany}}, \
  and\ \bibinfo {author} {\bibfnamefont {A.}~\bibnamefont {Palyi}},\ }\href
  {\doibase https://doi.org/10.1007/978-3-319-25607-8} {\emph {\bibinfo {title}
  {A Short Course on Topological Insulators}}},\ \bibinfo {series} {Lecture
  Notes in Physics}, Vol.\ \bibinfo {volume} {919}\ (\bibinfo  {publisher}
  {Springer International Publishing},\ \bibinfo {year} {2016})\BibitemShut
  {NoStop}%
\end{thebibliography}
%

\end{document}